\documentclass[authoryear,final,times,5p,twocolumn]{elsarticle}

\usepackage{amssymb}
\usepackage{amsthm}

\usepackage{lineno}

\usepackage{color}

\journal{Icarus}

\begin{document}

\begin{frontmatter}



\title{Thermal conductivity and coordination number of compressed dust aggregates}


\author[label1]{Sota Arakawa}
\ead{arakawa.s.ac@m.titech.ac.jp}
\author[label2,label3]{Misako Tatsuuma}
\author[label4]{Naoya Sakatani}
\author[label1]{Taishi Nakamoto}

\address[label1]{Department of Earth and Planetary Sciences, Tokyo Institute of Technology, Meguro, Tokyo 152-8551, Japan}
\address[label2]{Department of Astronomy, Graduate School of Science, The University of Tokyo, Bunkyo, Tokyo 113-0033, Japan}
\address[label3]{Division of Theoretical Astronomy, National Astronomical Observatory of Japan, Mitaka, Tokyo 181-8588, Japan}
\address[label4]{Institute of Space and Astronautical Science, Japan Aerospace Exploration Agency, Sagamihara, Kanagawa 252-5210, Japan}

\begin{abstract}
Understanding the heat transfer mechanism within dust aggregates is of great importance for many subjects in planetary science.
We calculated the coordination number and the thermal conductivity through the solid network of compressed dust aggregates.
We found a simple relationship between the coordination number and the filling factor and revealed that the thermal conductivity through the solid network of aggregates is represented by a power-law function of the filling factor and the coordination number.
\end{abstract}


\end{frontmatter}


\section{Introduction}

Understanding the thermal conductivity of dust aggregates and powdered media is important in numerous scientific and engineering fields.
In the context of planetary sciences, for example, the thermal evolution of planetesimals is affected by the thermal conductivity of dust aggregates because they were formed from micron-sized grains in the solar nebula \citep[e.g.,][]{Henke+2013,Sirono2017}.
The near-surface temperature distribution of comets and asteroids also depends on the thermal conductivity of surface grains called regolith \citep[e.g.,][]{Blum+2017,Okada+2017}.
Moreover, the radial motion of dust aggregates in a protoplanetary disk is induced by photophoresis, and the efficiency of this process is 
controlled by the thermal conductivity \citep[e.g.,][]{Wurm+2009,Loesche+2012}.

The thermal conductivity of dust aggregates depends on many physical parameters, and there are many experimental and numerical studies on the thermal conductivity of dust aggregates.
The thermal conductivity of porous aggregates under a vacuum condition is given by two terms: the thermal conductivity through the solid network $k_{\rm sol}$ and the thermal conductivity owing to radiative transfer $k_{\rm rad}$.
It is usually thought that the coordination number of monomer grains (i.e., the average number of contacts per grain) $Z$ influences the thermal conductivity through the solid network $k_{\rm sol}$ \citep[e.g.,][]{Gusarov+2003,Sirono2014}.
\citet{Sakatani+2016} revealed that $k_{\rm sol}$ is dependent on the contact radius between monomers $r_{\rm c}$ normalized by the monomer radius $R$ as well.
The thermal conductivity owing to radiative transfer $k_{\rm rad}$ is affected by the temperature of dust aggregates $T$ and the mean free path of photons $l_{\rm p}$ \citep[e.g.,][]{Gundlach+2012,Arakawa+2017}, and the mean free path of photons $l_{\rm p}$ depends on the monomer radius $R$, the filling factor $\phi$, and optical properties of dust aggregates.

\citet{Arakawa+2017} revealed that the thermal conductivity through the solid network $k_{\rm sol}$ is proportional to the square of the filling factor $\phi$ for highly porous aggregates with filling factors below $10^{-1}$.
The coordination number $Z$ hardly changes with changing of $\phi$ for highly porous aggregates and the effect of $Z$ on $k_{\rm sol}$ would be invisible.
In contrast, for compressed aggregates with filling factors above $10^{-1}$, the coordination number $Z$ could be dependent on the filling factor $\phi$ and the effect of the coordination number on the thermal conductivity is expected to be observed.

In this study, we calculate the coordination number $Z$ and the thermal conductivity through the solid network $k_{\rm sol}$ for compressed dust aggregates with filling factors in a wide range of $\phi \lesssim 1$.
These snapshot data of compressed dust aggregates are prepared in the same way as \citet{Kataoka+2013a}.
We examine filling factor dependences of the coordination number $Z$ and the thermal conductivity through the solid network $k_{\rm sol}$, and we derive empirical formulae of $Z = Z {(\phi)}$ and $k_{\rm sol} = k_{\rm sol} {(\phi, Z {(\phi)})}$.
Then we confirm the validity of the results by comparison with the experimental data of \citet{Sakatani+2017}.
Our findings are expected to be competent tools for many fields of study related to dust aggregates and powdered media.

\section{Methods}

We perform three-dimensional numerical calculations of dust aggregates using the model described by \citet{Arakawa+2017}.
Here we briefly summarize our numerical methods.

\subsection{Arrangement of monomer grains}

\begin{figure}[h]
\centering
\includegraphics[width=0.9\columnwidth]{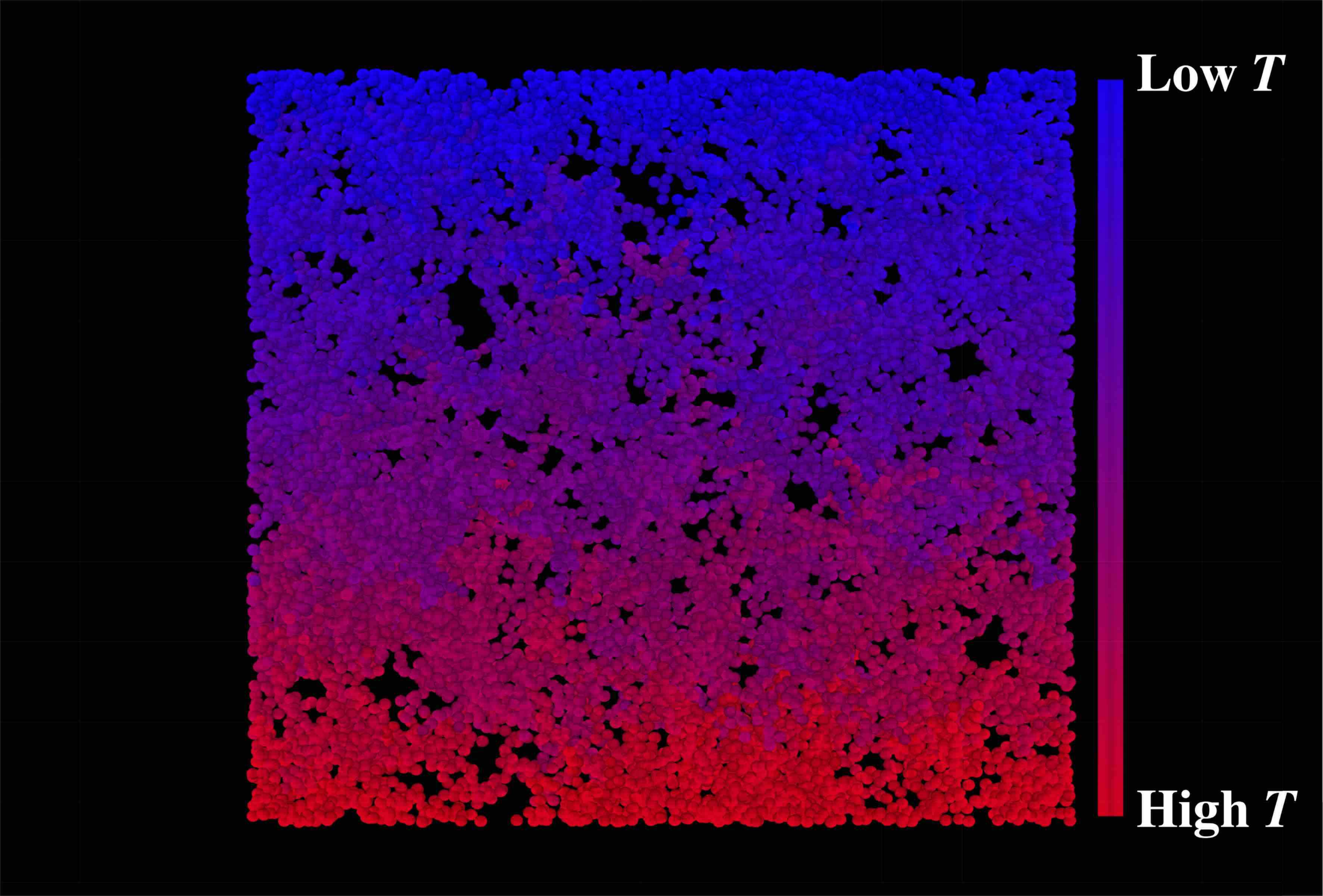}
\caption{
Example snapshot of a compressed dust aggregate.
The filling factor of the presented aggregate is $\phi = 10^{-1.5}$.
The colors of monomers, corresponding to the temperature of each monomer grains, are given by Eq.\ (\ref{eq1}).
}
\label{fig1}
\end{figure}

The arrangement of monomer grains depends on the coagulation history of the aggregates.
At the initial stage of coagulation of dust aggregates in protoplanetary disks, both experimental \citep[e.g.,][]{Wurm+1998} and theoretical \citep[e.g.,][]{Kempf+1999} studies have shown that hit-and-stick collisions without compression lead to the formation of highly porous aggregates with the fractal dimension close to two, which is called ballistic cluster-cluster aggregation \citep[BCCA;][]{Meakin1991}.
In this study, we prepare snapshots of the compressed BCCA aggregates comprised of $2^{14} = 16384$ spherical monomer grains using three-dimensional numerical simulations of static compression as \citet{Kataoka+2013a}.
Figure \ref{fig1} is an example snapshot of a compressed dust aggregate.

\subsection{Temperature structure of the dust aggregate}

\begin{figure}[h]
\centering
\includegraphics[width=0.9\columnwidth]{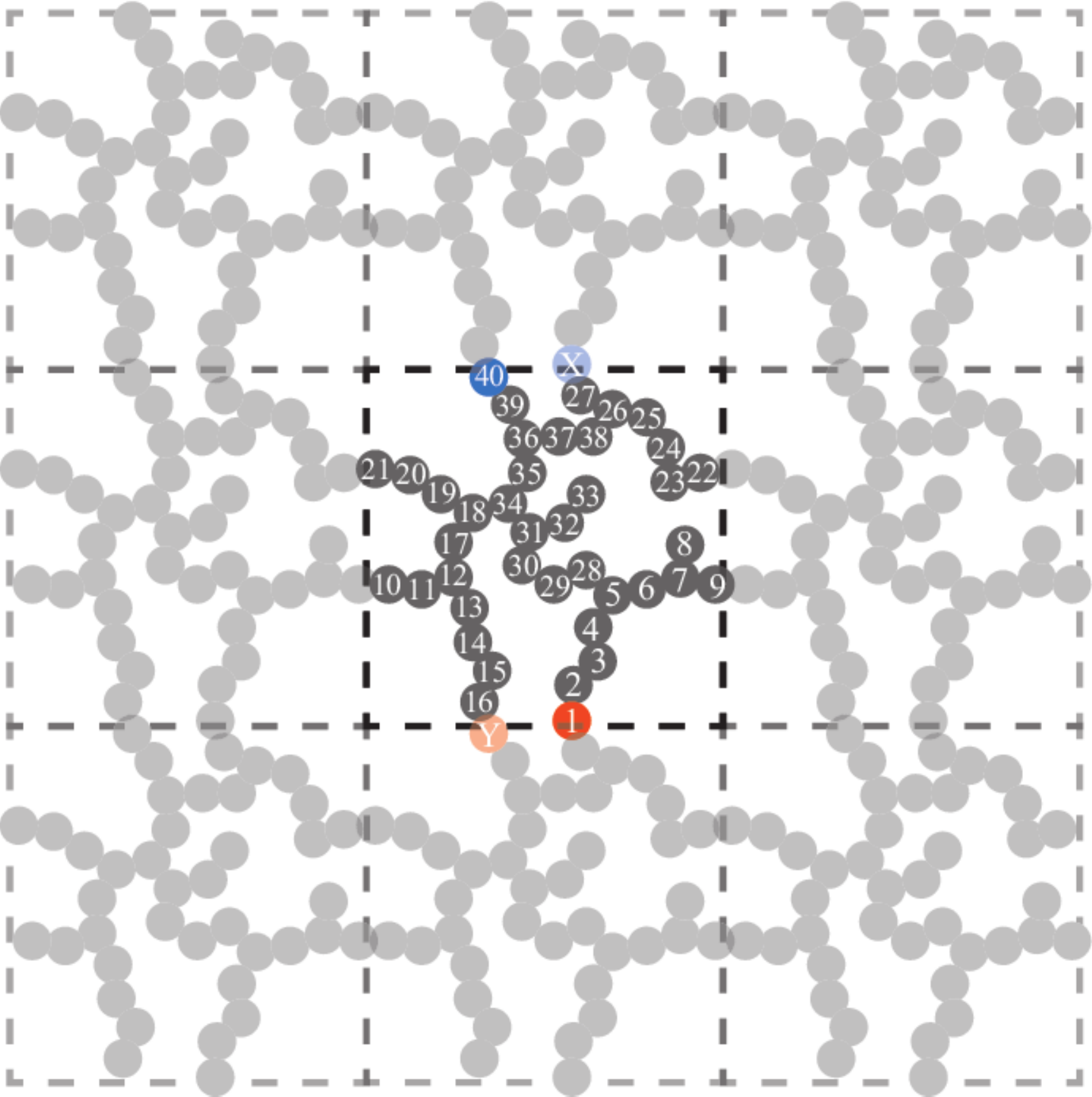}
\caption{
Sketch of a dust aggregate in a cubic periodic boundary.
The temperature of grains located on the lower (number 1 and location Y) and upper (number 40 and location X) boundary is set to $T_{0} + \Delta / 2$ and $T_{0} + \Delta / 2$, respectively.
The temperature of each grain is calculated by solving the equation of heat balance (Eq.\ \ref{eq1}) simultaneously for each grain \citep[from][]{Arakawa+2017}.
}
\label{fig2}
\end{figure}

In order to calculate the thermal conductivity through the solid network of an aggregate $k_{\rm sol}$, we determine the temperature of each monomer grain in a cubic periodic boundary \citep{Sirono2014,Arakawa+2017}.
We calculate the temperature of each grain using the method of \citet{Arakawa+2017}.
Here, we consider one-directional heat flow from the lower boundary plane to the upper boundary plane (see Fig.\ \ref{fig2}).
There are three choices regarding the pair of the lower and the upper planes, and we calculate $k_{\rm sol}$ from three directions.

We define $R$ as the monomer radius and $L^{3}$ as the volume of each cubic space.
The location of the $i$-th grain $(x_{i}, y_{i}, z_{i})$ satisfies $|x_{i}| < L/2$, $|y_{i}| < L/2$, and $|z_{i}| < L/2$ for $i = 1, 2, ..., N$, where $N = 16384$ is the number of grains in the periodic boundary (see Fig.\ \ref{fig2}).
The grains located in $- L/2 < z_{i} < - (L/2 - R)$ are on the lower boundary, and the grains located in $+ (L/2 - R) < z_{i} < + L/2$ are on the upper boundary.
When the $i$-th grain is located on the lower (upper) boundary, we add a new grain on the upper (lower) boundary.
The location of the new grain is $(x_{i}, y_{i}, z_{i} + L)$ for the case when the $i$-th grain is located on the lower boundary and $(x_{i}, y_{i}, z_{i} - L)$ for the case when the $i$-th grain is located on the upper boundary.
We set the temperature of grains located on the lower and the upper boundary as $T_{0} + {\Delta T}/2$ and $T_{0} - {\Delta T}/2$, respectively.

Heat flows through the monomer-monomer contacts, and for the case of the steady state, the equation of heat balance at the internal $i$-th grain is given by
\begin{equation}
\sum_{j} F_{i, j} = 0,
\label{eq1}
\end{equation}
where $F_{i, j}$ is the heat flow from the $j$-th grain to the $i$-th grain.
The heat flow from the $j$-th grain to the $i$-th grain $F_{i, j}$ is given by
\begin{equation}
F_{i, j} = H_{\rm c} {(T_{j} - T_{i})},
\end{equation}
where $H_{\rm c}$ is the heat conductance at the contact of two grains and $T_{i}$ and $T_{j}$ are the temperatures of the $i$-th and $j$-th grains.
We consider the contacts not only inside the periodic boundary but also on the side boundaries.
The heat conductance at the contact of two grains $H_{\rm c}$ is \citep{Cooper+1969}
\begin{equation}
H_{\rm c} = 2 k_{\rm mat} r_{\rm c},
\end{equation}
where $k_{\rm mat}$ is the material thermal conductivity and $r_{\rm c}$ is the contact radius of monomer grains.
The contact radius $r_{\rm c}$ depends on the monomer radius $R$ and the material parameters \citep[see, e.g.,][]{Wada+2007};
\begin{equation}
r_{\rm c} = {\left( \frac{9 \pi \gamma {(1 - {\nu_{\rm P}}^{2})}}{2 Y R} \right)}^{1/3} R,
\label{eqr}
\end{equation}
where $\gamma$, $\nu_{\rm P}$, and $Y$ are the surface energy, the Poisson's ratio, and Young's modulus of monomer grains, respectively.
The temperature structure of the aggregate in the cubic periodic boundary can be calculated by solving the Eq.\ (\ref{eq1})  simultaneously for all $N$ monomer grains, except lower and upper boundary grains, as shown in Fig.\ \ref{fig1}.

\subsection{Thermal conductivity through the solid network}

Once the temperature structure is obtained, we calculate the total heat flow at the upper boundary $\sum_{\rm upper} F_{i, j}$, where we take the sum of contacts between the upper boundary $i$-th grain and internal $j$-th grain (for the case of Fig.\ \ref{fig2}, $\sum_{\rm upper} F_{i, j} = F_{X, 27} + F_{40, 39}$).
The total heat flow at the upper boundary $\sum_{\rm upper} F_{i, j}$ can be rewritten using the thermal conductivity through the solid network $k_{\rm sol}$ as
\begin{equation}
\sum_{\rm upper} F_{i, j} = k_{\rm sol} \frac{\Delta T}{L} L^{2}.
\end{equation}
We discuss $k_{\rm sol}$ as a function of the filling factor $\phi$ in this study, and rewrite $L$ using $\phi$ as
\begin{equation}
L = {\left( \frac{4 \pi N}{3 \phi} \right)}^{1/3} R.
\end{equation}
Therefore we obtain $k_{\rm sol}$ as a function of $\phi$ as follows:
\begin{eqnarray}
k_{\rm sol} &=& \frac{1}{L {\Delta T}} {\sum_{\rm upper} F_{i, j}}, \nonumber \\
&=& 2 k_{\rm mat} \frac{r_{\rm c}}{R} \cdot {\left( \frac{3 \phi}{4 \pi N} \right)}^{1/3} {\sum_{\rm upper} \frac{T_{j} - T_{i}}{{\Delta T}}}, \nonumber \\
&\equiv& 2 k_{\rm mat} \frac{r_{\rm c}}{R} \cdot f {(\phi)},
\label{eqk}
\end{eqnarray}
where $f {(\phi)}$ is a dimensionless function of $\phi$.

\section{Numerical Results}

We carried out 10 runs of numerical simulations of static compression with different initial shape of the BCCA aggregate.
Then we took 20 snapshot data for each run of compression simulation.
The filling factors of compressed dust aggregates range from $\phi = 10^{-2}$ to $10^{-0.4}$ with logarithmic steps of $10^{0.1}$ and from $10^{-0.4}$ to $10^{-0.25}$ with logarithmic steps of $10^{0.05}$.
We investigate the filling factor dependence of $Z {(\phi)}$ and $f {(\phi)}$ from 10 snapshot data for each $\phi$ obtained from different runs of compression simulations.

\subsection{Coordination number}

The coordination number (i.e., the average number of contacts per grain) $Z$ increases as an aggregate is compressed.
The initial coordination number of an uncompressed BCCA aggregate is approximately $Z \simeq 2$, so we define $\zeta$ as the deviation of the coordination number from the initial condition;
\begin{equation}
\zeta \equiv Z - 2.
\end{equation}

\begin{figure}[h]
\centering
\includegraphics[width=\columnwidth]{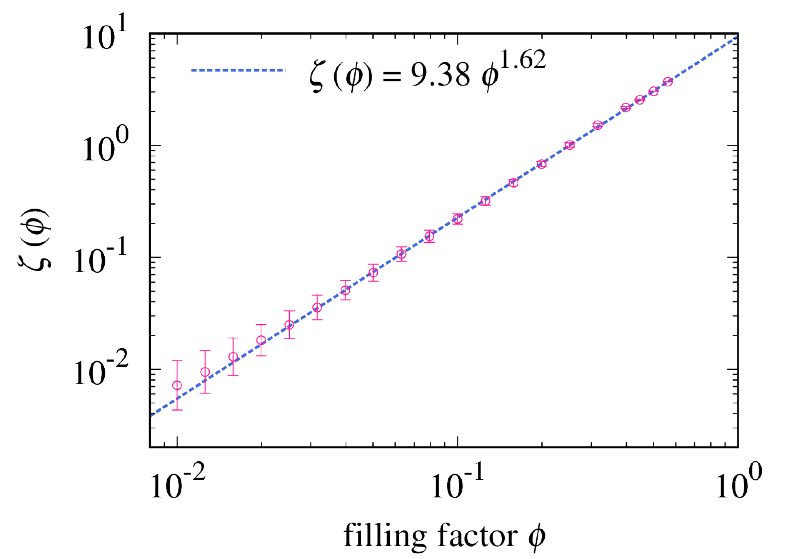}
\caption{
Fitting of the deviation of the coordination number from two $\zeta \equiv Z - 2$ as a function of the filling factor $\phi$.
The magenta circles represent the averaged data with vertical error bars of twice the standard error.
The blue dashed curve is the best-fit obtained from the weighted least-squares method.
}
\label{fig3}
\end{figure}

Figure \ref{fig3} shows the deviation of the coordination number from the initial condition $\zeta {(\phi)}$ as a function of the filling factor $\phi$.
The magenta circles represent the geometric mean of 10 snapshots from different runs with vertical error bars of twice the standard error.
We found that a power-law function well fits the filling factor dependence of $\zeta {(\phi)}$ at least in the range of $10^{-2} < \phi < 10^{-0.25}$, and the best-fit curve given by the weighted least-squares method is (blue dashed curve),
\begin{equation}
\zeta {(\phi)} = 9.38 \phi^{1.62}.
\end{equation}
Therefore the coordination number $Z {(\phi)}$ is given by,
\begin{equation}
Z {(\phi)} = 2 + 9.38 \phi^{1.62}.
\label{eqz}
\end{equation}

The coordination number and the filling factor are the key parameters not only for the thermal conductivity but also for the outcome of the collision between dust aggregates \citep[e.g.,][]{Wada+2011,Seizinger+2013}.
Bouncing collisions of dust aggregates within a protoplanetary disk might prevent dust aggregates from growing into planetesimals via direct aggregation \citep{Zsom+2010}.
The results from numerical simulations of aggregate collisions indicate that dust aggregates can stick to each other only when the coordination number is $Z \lesssim 6$ \citep{Wada+2011}, or a filling factor of $\phi \lesssim 0.5$ might be the condition for collisional growth \citep{Seizinger+2013}.

\subsection{Thermal conductivity}

Previous studies \citep[e.g.,][]{Sakatani+2017} predicted that thermal conductivity depends on both the filling factor $\phi$ and the coordination number $Z$.
For highly porous aggregates, however, $Z {(\phi)}$ is approximately two and behaves as a constant, hence $f {(\phi)}$ would only depend on $\phi$ for highly porous aggregates.
\citet{Arakawa+2017} revealed that $f {(\phi)}$ is approximately proportional to the square of $\phi$ for highly porous aggregates with filling factors below $10^{-1}$.
In this study, we calculate the dimensionless function of the thermal conductivity $f {(\phi)}$ for both loose (i.e., $Z \simeq 2$) and close (i.e., $Z > 2$) dust aggregates.

\begin{figure}[h]
\centering
\includegraphics[width=\columnwidth]{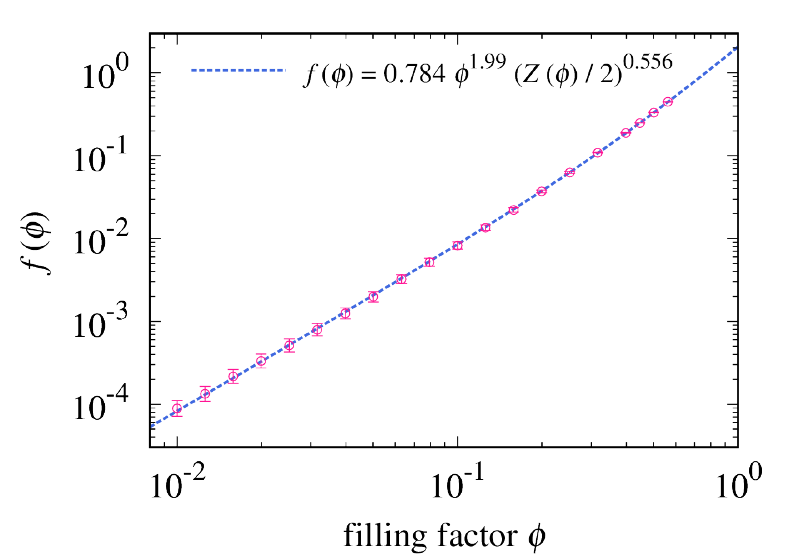}
\caption{
Fitting of the dimensionless function of thermal conductivity $f {(\phi)}$ as a function of the filling factor $\phi$.
The magenta circles represent the averaged data with vertical error bars of twice the standard error.
The blue dashed curve is the best-fit obtained from the weighted least-squares method.
}
\label{fig4}
\end{figure}

Figure \ref{fig4} shows the dimensionless function $f {(\phi)}$ as a function of the filling factor $\phi$.
The magenta circles represent the geometric mean of 30 calculation results of the temperature structure from 3 directions and 10 different runs, with vertical error bars of twice the standard error.
Here we assume that $f {(\phi)}$ is given by the power-law function of $\phi$ and $Z {(\phi)}$, and the best-fit curve given by the weighted least-squares method is (blue dashed curve),
\begin{equation}
f {(\phi)} = 0.784 \phi^{1.99} {\left( \frac{Z {(\phi)}}{2} \right)}^{0.556}.
\label{eqf}
\end{equation}

For highly porous dust aggregates with filling factors in the range from $10^{-2}$ to $10^{-1}$, \citet{Arakawa+2017} obtained a relationship between $f {(\phi)}$ and $\phi$ as $f {(\phi)} \simeq \phi^{2}$.
Our novel formula of $f {(\phi)}$ approximately coincides with the result of \citet{Arakawa+2017} for highly porous aggregates, and we numerically reveal the effect of the coordination number $Z$ on the dimensionless function of the thermal conductivity $f$ using highly compressed aggregates.

\section{Comparison with experimental data}

\subsection{Comparison with \citet{Sakatani+2017}}

\begin{figure}[h]
\centering
\includegraphics[width=\columnwidth]{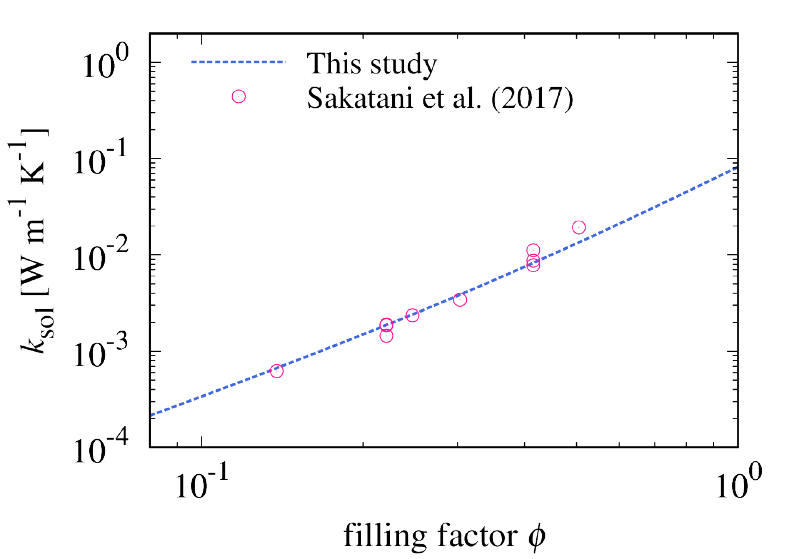}
\caption{
The thermal conductivity through the solid network $k_{\rm sol}$ of dust aggregates composed of ${\rm Si}{\rm O}_{2}$ glass grains.
We compare experimental data of \citet[][magenta circles]{Sakatani+2017} with our empirical model (blue dashed curve).
The monomer radius is $R = 2.5\ \mu{\rm m}$ and the material thermal conductivity $k_{\rm mat}$ at a temperature of $300\ {\rm K}$ is $k_{\rm mat} = 1.57\ {\rm W}\ {\rm m}^{-1}\ {\rm K}^{-1}$ \citep{Sakatani+2017}.
}
\label{fig5}
\end{figure}

The thermal conductivity of dust aggregates in a vacuum is the sum of $k_{\rm sol}$ and $k_{\rm rad}$, and the contributions of $k_{\rm sol}$ and $k_{\rm rad}$ are distinguishable by measuring the temperature dependence of the total thermal conductivity \citep[e.g.,][]{Sakatani+2016,Sakatani+2017}.
\citet{Sakatani+2017} obtained the filling factor dependence of $k_{\rm sol}$ for dust aggregates composed of micron-sized glass grains.
Therefore, by comparison with the experimental data of \citet{Sakatani+2017}, we can verify our model (Fig.\ \ref{fig5}).
The blue dashed curve is the calculated thermal conductivity from Eqs.\ (\ref{eqk}) and (\ref{eqf}), and the magenta circles represents the experimental data of dust aggregates \citep{Sakatani+2017}.

When we consider the dust aggregates of micron-sized monomers, the contact radius between monomers $r_{\rm c}$ is given by Eq.\ (\ref{eqr}).
The material properties of ${\rm Si}{\rm O}_{2}$ glass grains are listed as follows: the surface energy $\gamma = 20\ {\rm mJ}\ {\rm m}^{-2}$, the Poisson's ratio $\nu_{\rm P} = 0.17$, and Young's modulus $Y = 54\ {\rm GPa}$ \citep{Seizinger+2013}.
The monomer radius used in \citet{Sakatani+2017} is $R = 2.5\ \mu{\rm m}$.
The material thermal conductivity $k_{\rm mat}$ depends on the temperature, and at a temperature of $300\ {\rm K}$, the material thermal conductivity is $k_{\rm mat} = 1.57\ {\rm W}\ {\rm m}^{-1}\ {\rm K}^{-1}$ \citep{Sakatani+2017}.
Figure \ref{fig5} shows that our empirical model of $k_{\rm sol}$ reproduces the experimental results with a relative difference of 30\% or less for a wide range of filling factors.

\subsection{Comparison with \citet{Krause+2011}}

The thermal conductivity of porous dust aggregates composed of micron-sized ${\rm Si}{\rm O}_{2}$ glass grains is obtained by \citet{Krause+2011}, using a combination of laboratory experiments and numerical simulations.
\citet{Krause+2011} reported the filling factor dependence of the thermal conductivity of dust aggregates as \citet{Sakatani+2017}, although \citet{Krause+2011} did not resolve the contributions of the thermal conductivity through the solid network $k_{\rm sol}$ and the thermal conductivity owing to radiative transfer $k_{\rm rad}$.
For dust aggregates composed of micron-sized monomers, however, we can calculate $k_{\rm rad}$ using the Rossland diffusion approximation \citep[e.g.,][]{Arakawa+2017}, and we can also calculate $k_{\rm sol}$ by using Eqs.\ (\ref{eqk}) and (\ref{eqf}).
Therefore, by comparison with the experimental data of \citet{Krause+2011}, we can test the validity of our novel models for both $k_{\rm sol}$ and $k_{\rm rad}$.

The thermal conductivity through the solid network $k_{\rm sol}$ is calculated from Eqs.\ (\ref{eqk}) and (\ref{eqf}).
In the experiments by \citet{Krause+2011}, each sample was heated by the laser beam, and the temperature of dust aggregates temporally changes during their thermal conductivity measurements.
\citet{Krause+2011} reported that the surface temperature of heated dust samples varied from $300\ {\rm K}$ to $450\ {\rm K}$. 
Therefore we take into account the temperature dependence of the contact radius $r_{\rm c}$ and the material thermal conductivity $k_{\rm mat}$. 
We consider the temperature dependence of the material thermal conductivity $k_{\rm mat}$ and the surface energy $\gamma$.
\citet{Gundlach+2012} assumed that $k_{\rm mat}$ and $\gamma$ increase with temperature as follows:
\begin{equation}
k_{\rm mat} = c_{1} +  c_{2} T,
\end{equation}
and
\begin{equation}
\gamma = c_{3} T,
\end{equation}
with $c_{1} = 9.94 \times 10^{-1}\ {\rm W}\ {\rm K}^{-1}\ {\rm m}^{-1}$, $c_{2} = 1.26 \times 10^{-3}\ {\rm W}\ {\rm K}^{-2}\ {\rm m}^{-1}$, and $c_{3} = 6.67 \times 10^{-5}\ {\rm J}\ {\rm m}^{-2}\ {\rm K}^{-1}$.
We apply these temperature dependences of $k_{\rm mat}$ and $\gamma$ to the calculation of $k_{\rm sol}$.
The monomer radius used in \citet{Krause+2011} is $R = 0.75\ \mu{\rm m}$, and we assume the Poisson's ratio $\nu_{\rm P} = 0.17$ and Young's modulus $Y = 54\ {\rm GPa}$ \citep{Seizinger+2013} as constant values for the calculation of the contact radius $r_{\rm c}$.

We also calculate the thermal conductivity owing to radiative transfer $k_{\rm rad}$.
The thermal conductivity owing to radiative transfer is proportional to the mean free path of photons $l_{\rm p}$.
In addition, when the monomer radius is smaller than the thermal emission wavelength, $l_{\rm p}$ is inversely proportional to the Rossland mean opacity $\kappa_{\rm R}$.
Therefore we calculate the Rossland mean opacity of ${\rm Si}{\rm O}_{2}$ glass grains whose radius is $R = 0.75\ \mu{\rm m}$ and we obtain the temperature-dependent $\kappa_{\rm R}$ (see \ref{appA}).

\begin{figure}[h]
\centering
\includegraphics[width=\columnwidth]{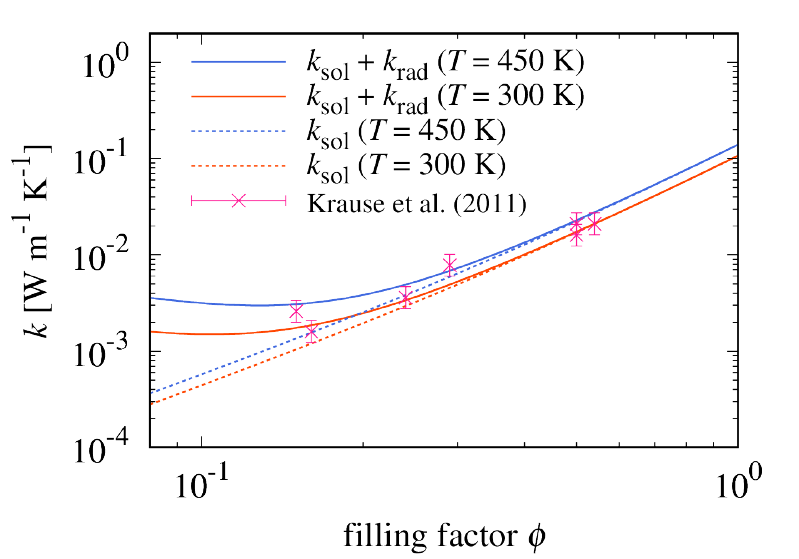}
\caption{
The thermal conductivity of dust aggregates composed of ${\rm Si}{\rm O}_{2}$ glass grains.
We compare the experimental data of \citet[][magenta crosses with $30\%$ error bars]{Krause+2011} with our empirical model (blue and red curves).
The solid curves represent the sum of the thermal conductivity through the solid network and the thermal conductivity owing to radiative transfer, $k_{\rm sol} + k_{\rm rad}$, whereas the dashed curves represent the thermal conductivity through the solid network $k_{\rm sol}$.
The blue curves correspond to the thermal conductivity at $T = 450\ {\rm K}$, and the red curves correspond to the thermal conductivity at $T = 300\ {\rm K}$.
The monomer radius is $R = 0.75\ \mu{\rm m}$, and the temperature-dependence of the material thermal conductivity $k_{\rm mat}$ and the surface energy $\gamma$ is considered \citep{Gundlach+2012}.
The thermal conductivity owing to radiative transfer $k_{\rm rad}$ is calculated from Eqs.\ (\ref{eqkrad}) and (\ref{eqlp}).
}
\label{fig6}
\end{figure}

Figure \ref{fig6} shows the filling factor dependence of the thermal conductivity through of dust aggregates composed of ${\rm Si}{\rm O}_{2}$ glass grains.
The solid curves are the sum of the thermal conductivity through the solid network and the thermal conductivity owing to radiative transfer, $k_{\rm sol} + k_{\rm rad}$, whereas the dashed curves represent the thermal conductivity through the solid network $k_{\rm sol}$.
The blue curves correspond to the thermal conductivity at $T = 450\ {\rm K}$, and the red curves correspond to the thermal conductivity at $T = 300\ {\rm K}$.
By comparison with experimental data of \citet[][magenta crosses]{Krause+2011}  and our numerical result of $k_{\rm sol} + k_{\rm rad}$ (blue and orange solid curves), we found that our novel model well reproduces the experimental results for a wide range of filling factors.
In addition, the experimental results can only be explained when we take into consideration the effect of radiative transfer when we consider highly porous aggregates (see the experimental data of $\phi = 0.15$ in Fig.\ \ref{fig6}).
This conclusion about the contribution of $k_{\rm rad}$ is qualitatively consistent with the claim of \citet{Gundlach+2012}.
We also stress that our model has no free parameter to fit the numerical calculations to the experimental results.

\section{Discussion}

\subsection{Contributions of $k_{\rm sol}$ and $k_{\rm rad}$}

As shown in Fig.\ \ref{fig6}, the contribution of $k_{\rm rad}$ becomes important when the filling factor of dust aggregates becomes lower, whereas the contribution of $k_{\rm sol}$ is much important when the filling factor is higher.
Here we define the transition filling factor $\phi_{\rm sol-rad}$ by solving the equation $k_{\rm sol} = k_{\rm rad}$.
Assuming $f {(\phi)} \simeq \phi^{2}$ for simplicity, we obtain the following relation:
\begin{equation}
\phi_{\rm sol-rad} \simeq {\left( \frac{8 \sigma_{\rm SB}}{3 \rho_{\rm mat}} \frac{R}{r_{\rm c}} \frac{1}{k_{\rm mat} \kappa_{\rm R}} \right)}^{1/3} T.
\end{equation}
For the case of ${\rm Si}{\rm O}_{2}$ glass grains with $T = 300\ {\rm K}$ and $R = 0.75\ {\mu}{\rm m}$, the transition filling factor $\phi_{\rm sol-rad}$ is given by $\phi_{\rm sol-rad} = 0.13$.
The transition filling factor $\phi_{\rm sol-rad}$ is a strong function of $T$ because of the strong temperature dependence of $k_{\rm rad}$; when the temperature is $T \ll 100\ {\rm K}$, then the transition filling factor is $\phi_{\rm sol-rad} \ll 10^{-1}$ for dust aggregates composed of micron-sized ${\rm Si}{\rm O}_{2}$ glass grains.
This fact implies that the thermal conductivity through the solid network $k_{\rm sol}$ might be the dominant term for the heat conduction within cold small bodies, such as comets and trans-Neptunian objects.

\subsection{Estimation of monomer properties}

As seen above, both $k_{\rm sol}$ and $k_{\rm rad}$ depend on the chemical composition and the monomer radius of dust aggregates, because $k_{\rm mat}$, $\kappa_{\rm R}$, and $r_{\rm c} / R$ are strongly dependent on these monomer properties.
Therefore, it is possible to give some constraints on the chemical composition and the monomer radius of small bodies by measuring the thermal conductivity of the surface regolith of small bodies.
Determining these fundamental properties is of course exceedingly important for understanding the collisional growth process of the planetesimal formation \citep[e.g.,][]{Blum+2008,Arakawa+2016b,Musiolik+2016b}.
For the case of comets, the thermal inertia of comets 9P/Tempel 1 and 67P/Churyumov-Gerasimenko are observed by the {\it Deep Impact} mission and the {\it Rosetta} mission \citep[e.g.][]{Davidsson+2013,Spohn+2015,Marshall+2018}.
The thermal inertia $I$ is given by $I = \sqrt{{(k_{\rm sol} + k_{\rm rad})} \rho_{\rm mat} C_{\rm heat} \phi}$, where $C_{\rm heat}$ is the specific heat.
Then we can evaluate the ${\left( k_{\rm sol} + k_{\rm rad} \right)} \rho_{\rm mat} C_{\rm heat}$ when $I$ and $\phi$ are known from observations, and the product ${\left( k_{\rm sol} + k_{\rm rad} \right)} \rho_{\rm mat} C_{\rm heat}$ is a parameter directly related to the chemical composition and the monomer radius.
We will discuss the nature of the building block of comets by using thermal conductivity calculations.

\subsection{Pebble-pile hypothesis}
In this study, we consider the heat conduction within homogeneous dust aggregates.
However, several studies \citep[e.g.,][]{Blum+2017,WahlbergJansson+2017} proposed that comets were formed via collapse of gravitationally bound clouds of millimeter- to centimeter-sized compactified dust aggregates so-called pebbles, and pebble clouds might collapse into porous pebble-pile bodies.
For pebble-pile bodies comprised of millimeter- to centimeter-sized pebbles, the thermal conductivity owing to radiative transfer would be the dominant term for the heat conduction due to the large mean free path of photons inside the voids between pebbles \citep{Gundlach+2012}.
We will address the issue of ``pebble-pile hypothesis'' in future.

\section{Conclusion}

In this study, we conducted numerical simulations to determine the filling factor dependence of the thermal conductivity of compressed dust aggregates.
The initial arrangements of aggregates are given by BCCA and numerical simulations of static compression are done in a cubic periodic boundary.
We found a simple relationship between the coordination number $Z$ and the filling factor $\phi$, $Z {(\phi)} = 2 + 9.38 \phi^{1.62}$.
This relationship is practical in a wide range of filling factors.
It is also revealed that the thermal conductivity through the solid network $k_{\rm sol}$ is given by a power-law function of the filling factor $\phi$ and the coordination number $Z {(\phi)}$ as $k_{\rm sol} \propto \phi^{1.99} {Z {(\phi)}}^{0.556}$.
Although what these indices come from is still unclear, this empirical formula well explains the experimental data of the thermal conductivity measured by \citet{Sakatani+2017}.
In addition, when we consider the contributions of both the thermal conductivity through the solid network and the thermal conductivity owing to radiative transfer, our novel model can reproduce the experimental results of \citet{Krause+2011} without any free parameters to fit the numerical calculations to the experimental results.

By using our novel relations, we can determine the thermal conductivity of dust aggregates from a given set of physical parameters.
Conversely, we can also constrain the physical parameters of a specific sample by measuring the thermal conductivity.
Therefore, our findings are expected to be competent tools for many fields of study related to dust aggregates and powdered media.

\section*{Acknowledgements}

We thank Sin-iti Sirono and Masaki Takemoto for fruitful discussions.
S.A.\ acknowledges Kazumasa Ohno for providing the extinction cross section of ${\rm Si}{\rm O}_{2}$ amorphous grains.
This work is supported by JSPS KAKENHI Grant (JP18K03721).
S.A.\ is supported by the Grant-in-Aid for JSPS Research Fellow (JP17J06861).
N.S.\ is partly supported by JSPS Grant-in-Aid for Scientific Research on Innovative Areas (JP17H06459) and JSPS Core-to-Core Program ``International Network of Planetary Sciences''.

\appendix

\section{Thermal conductivity owing to radiative transfer}
\label{appA}

In this study, we consider the thermal conductivity owing to radiative transfer within fluffy aggregates composed of micron-sized grains.
The thermal conductivity owing to radiative transfer $k_{\rm rad}$ is given by \citep[e.g.,][]{Merrill1969}
\begin{equation}
k_{\rm rad} = \frac{16}{3} \sigma_{\rm SB} T^{3} l_{\rm p},
\label{eqkrad}
\end{equation}
where $\sigma_{\rm SB} = 5.67 \times 10^{-8}\ {\rm W}\ {\rm m}^{-2}\ {\rm K}^{-4}$ is the Stefan-Boltzmann constant.
We calculated the mean free path of photons in fluffy aggregates of micron-sized grains $l_{\rm p}$ as follows:
\begin{equation}
l_{\rm p} = \frac{1}{\kappa_{\rm R} \rho_{\rm mat} \phi},
\label{eqlp}
\end{equation}
where $\kappa_{\rm R}$ is the Rossland mean opacity and $\rho_{\rm mat}$ is the material density.
We also note that when the wavelength of the thermal radiation is shorter than the monomer radius and monomer grains are considered as opaque, we can apply the geometrical optics approximation for the evaluation of $l_{\rm p}$ \citep[e.g.,][]{Schotte1960}.

\begin{figure}[h]
\centering
\includegraphics[width=\columnwidth]{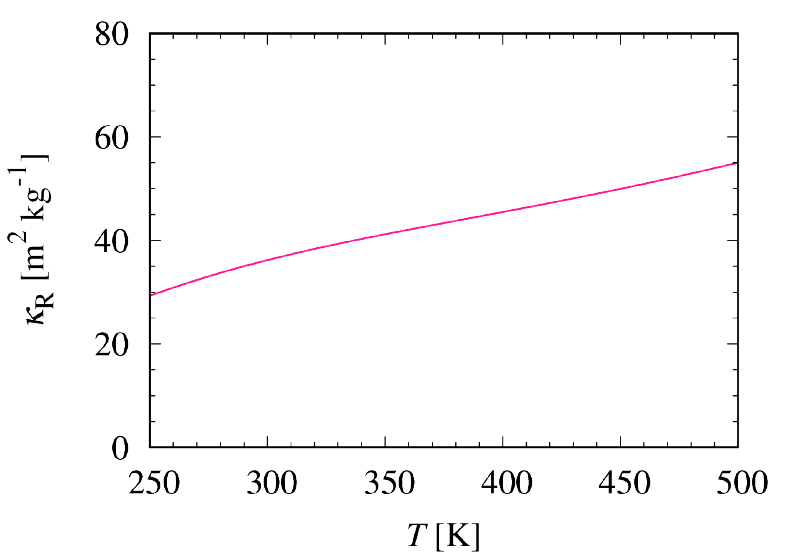}
\caption{
The Rossland mean opacity of ${\rm Si}{\rm O}_{2}$ glass grains.
The monomer radius is $R = 0.75\ {\mu}{\rm m}$ and the material density is $\rho_{\rm mat} = 2.0 \times 10^{3}\ {\rm kg}\ {\rm m}^{-3}$.
The refractive index of ${\rm Si}{\rm O}_{2}$ amorphous grains is taken from \citet{Kitzmann+2018}.
}
\label{fig7}
\end{figure}

The wavelength-averaged mean free path of photons $l_{\rm p}$ is given by a mean opacity averaged over all wavelengths called the Rossland mean opacity $\kappa_{\rm R}$: 
\begin{equation}
\frac{1}{\kappa_{\rm R}} \equiv \frac{\int {\rm d}\nu\ {\kappa_{\rm ext}}^{-1} {\left( {{\partial}B_{\nu}}/{{\partial}T} \right)}}{\int {\rm d}\nu\ {\left( {{\partial}B_{\nu}}/{{\partial}T} \right)}},
\label{eqkappa}
\end{equation}
where $\nu$ is the frequency of photons, $\kappa_{\rm ext}$ is the wavelength-dependent extinction opacity, and $B_{\nu}$ is the Planck function.
The frequency $\nu$ can be rewritten as $\nu = c / \lambda$, where $c = 3.00 \times 10^{8}\ {\rm m}\ {\rm s}^{-1}$ is the speed of light and $\lambda$ is the wavelength.
We integrated Eq.\ (\ref{eqkappa}) from $\lambda = 0.1\ \mu{\rm m}$ to $100\ \mu{\rm m}$. 
The Planck function $B_{\nu}$ is given by,
\begin{equation}
B_{\nu} = \frac{2 h {\nu}^{3}}{c^{2}} \frac{1}{\exp{\left[ {(h \nu)} / {(k_{\rm B} T)} \right]} - 1},
\end{equation}
where $h = 6.63 \times 10^{-34}\ {\rm J}\ {\rm s}$ is the Planck constant and $k_{\rm B} = 1.38 \times 10^{-23}\ {\rm J}\ {\rm K}^{-1}$ is the Boltzmann constant, respectively.
With the assumption of spherical homogeneous particles, the extinction cross section $C_{\rm ext}$ and opacity $\kappa_{\rm ext}$ are given by the Mie theory \citep{Mie1908,Bohren+1983}:
\begin{equation}
C_{\rm ext} = \frac{2 \pi R^{2}}{x^{2}} \sum_{n = 1}^{\infty} {\left( 2 n + 1 \right)} {\rm Re} {\left( a_{n} + b_{n} \right)},
\end{equation}
and
\begin{equation}
\kappa_{\rm ext} = \frac{C_{\rm ext}}{{\left( 4 \pi / 3 \right)} R^{3} \rho_{\rm mat}},
\end{equation}
where $x = 2 \pi R / \lambda$ is called the size parameter, and $a_{n}$ and $b_{n}$ are the Lorenz-Mie coefficients.
Then we can calculate the Lorenz-Mie coefficients from the size parameter $x$ and the wavelength-dependent complex refractive index $m$ from  the size parameter $x$ and the wavelength-dependent complex refractive index $m$ \citep[see, e.g.,][for details]{Bohren+1983}.
We took the wavelength-dependent complex refractive index of ${\rm Si}{\rm O}_{2}$ amorphous grains from \citet{Kitzmann+2018}.
The calculated Rossland mean opacity of ${\rm Si}{\rm O}_{2}$ glass grains is shown in Fig.\ \ref{fig7}.

  \bibliographystyle{elsarticle-harv} 

\newcommand*\aap{A\&A}
\let\astap=\aap
\newcommand*\aapr{A\&A~Rev.}
\newcommand*\aaps{A\&AS}
\newcommand*\actaa{Acta Astron.}
\newcommand*\aj{AJ}
\newcommand*\ao{Appl.~Opt.}
\let\applopt\ao
\newcommand*\apj{ApJ}
\newcommand*\apjl{ApJ}
\let\apjlett\apjl
\newcommand*\apjs{ApJS}
\let\apjsupp\apjs
\newcommand*\aplett{Astrophys.~Lett.}
\newcommand*\apspr{Astrophys.~Space~Phys.~Res.}
\newcommand*\apss{Ap\&SS}
\newcommand*\araa{ARA\&A}
\newcommand*\azh{AZh}
\newcommand*\baas{BAAS}
\newcommand*\bac{Bull. astr. Inst. Czechosl.}
\newcommand*\bain{Bull.~Astron.~Inst.~Netherlands}
\newcommand*\caa{Chinese Astron. Astrophys.}
\newcommand*\cjaa{Chinese J. Astron. Astrophys.}
\newcommand*\fcp{Fund.~Cosmic~Phys.}
\newcommand*\gca{Geochim.~Cosmochim.~Acta}
\newcommand*\grl{Geophys.~Res.~Lett.}
\newcommand*\iaucirc{IAU~Circ.}
\newcommand*\icarus{Icarus}
\newcommand*\jcap{J. Cosmology Astropart. Phys.}
\newcommand*\jcp{J.~Chem.~Phys.}
\newcommand*\jgr{J.~Geophys.~Res.}
\newcommand*\jqsrt{J.~Quant.~Spectr.~Rad.~Transf.}
\newcommand*\jrasc{JRASC}
\newcommand*\memras{MmRAS}
\newcommand*\memsai{Mem.~Soc.~Astron.~Italiana}
\newcommand*\mnras{MNRAS}
\newcommand*\na{New A}
\newcommand*\nar{New A Rev.}
\newcommand*\nat{Nature}
\newcommand*\nphysa{Nucl.~Phys.~A}
\newcommand*\pasa{PASA}
\newcommand*\pasj{PASJ}
\newcommand*\pasp{PASP}
\newcommand*\physrep{Phys.~Rep.}
\newcommand*\physscr{Phys.~Scr}
\newcommand*\planss{Planet.~Space~Sci.}
\newcommand*\pra{Phys.~Rev.~A}
\newcommand*\prb{Phys.~Rev.~B}
\newcommand*\prc{Phys.~Rev.~C}
\newcommand*\prd{Phys.~Rev.~D}
\newcommand*\pre{Phys.~Rev.~E}
\newcommand*\prl{Phys.~Rev.~Lett.}
\newcommand*\procspie{Proc.~SPIE}
\newcommand*\qjras{QJRAS}
\newcommand*\rmxaa{Rev. Mexicana Astron. Astrofis.}
\newcommand*\skytel{S\&T}
\newcommand*\solphys{Sol.~Phys.}
\newcommand*\sovast{Soviet~Ast.}
\newcommand*\ssr{Space~Sci.~Rev.}
\newcommand*\zap{ZAp}



\end{document}